\documentclass[amsmath,amssymb,aps,prl,twocolumn]{revtex4-2}
\usepackage{txfonts}
\usepackage{bm}
\usepackage{graphicx}

\newcommand{\be}{\begin{eqnarray}}
\newcommand{\ee}{\end{eqnarray}}
\newcommand{\bmat}{\left(\begin{array}}
\newcommand{\emat}{\end{array}\right)}
\newcommand{\no}{\nonumber}
\newcommand{\diff}{\mathrm d}
\newcommand{\e}{\mathrm e}

\begin{document}

\title{Performance Evaluation of Adiabatic Quantum Computation via Quantum Speed Limits \\ 
and Possible Applications to Many-Body Systems}
\author{Keisuke Suzuki}
\affiliation{Department of Physics, Tokyo Institute of Technology, Tokyo 152--8551, Japan}
\author{Kazutaka Takahashi}
\affiliation{Institute of Innovative Research, Tokyo Institute of Technology, Kanagawa 226--8503, Japan}

\date{\today}

\begin{abstract}
The quantum speed limit specifies a universal bound of the fidelity
between the initial state and the time-evolved state.
We apply this method to find a bound of the fidelity between
the adiabatic state and the time-evolved state.
The bound is characterized by the counterdiabatic Hamiltonian and 
can be used to evaluate the worst case performance 
of the adiabatic quantum computation.
The result is improved by imposing additional conditions and 
we examine several models to find a tight bound.
We also derive a different type of quantum speed limits
that is meaningful even when we take the thermodynamic limit.
By using solvable spin models, we study
how the performance and the bound are affected by phase transitions.
\end{abstract}

\maketitle

{\it Introduction.}
Knowing the fundamental speed limit for a dynamical process 
is an important problem in physics and 
is relevant to a broad range of research fields. 
Recent advances of quantum control technologies allow us
to discuss the fundamental limit even from a practical point of view.
In closed quantum systems, we can derive several limits 
known as quantum speed limits (QSLs)~\cite{MT,F,B,V,ML}.

Among many possible 
applications~\cite{Lloyd00,Lloyd02,LT,TEDdMF,dCEPH,DL,CYPCOD,DC,BSP}, 
we focus our attention on adiabatic quantum computation (AQC).
It is a method solving combinatorial optimization problems 
and has attracted intensive attention recently, 
as quantum annealing~\cite{KN,BBRA,FGGS,FGGLLP,AL18},
due to its use for a device manufactured 
by D-Wave Systems, Inc.~\cite{Jetal,BRIWWLMT}.
The solution of the problem is set to 
the ground state of the Ising Hamiltonian.
The Hamiltonian is slowly changed from 
a trivial form, represented by the transverse-field term, 
to the Ising Hamiltonian.
If the rate of the Hamiltonian change is very small, 
the time-evolved state is close to the instantaneous ground state
of the Hamiltonian.

The principle of AQC is based on the adiabatic theorem.
The infinitely slow time evolution is not realistic 
and we find nonadiabatic transitions in experiments.
Therefore, estimating and suppressing errors are important 
not only for its general use but also 
for understanding the dynamical properties of the quantum systems.
In the method of quantum adiabatic brachistochrone, 
a cost function is defined based on the notion of
adiabaticity~\cite{RKHLZ,RALZ,Takahashi19}.
It is minimized with respect to the protocol
to obtain an optimized algorithm.
The method practically gives a good performance 
but the result is strongly dependent on the choice of the cost function 
and does not assess the quantitative performance.
Some rigorous treatment of the adiabatic theorem allows us to derive
bounds of the performance~\cite{JRS,LRH,BS,EH},
but mathematically involved approaches are required 
for the derivation and the result is of limited use because of
the complicated expressions of the bound.

To overcome these problems, 
we employ the theory of QSL to find a rigorous bound. 
The QSL has a geometrical meaning and we can find a universal bound 
by using this approach.
Since the standard QSL only requires several simple inequalities,
the derivation is simple and 
the result can be written in an intuitively understandable form.
The QSL can give us a tight bound, which is also useful 
for practical applications such as the AQC.
We can further use the bound as a cost function to optimize the AQC.

We treat closed quantum systems throughout this paper.
For a given Hamiltonian $H(t)$ and a initial state $|\psi(0)\rangle$, 
unitary Schr\"odinger dynamics yields 
a time-evolved state $|\psi(t)\rangle$.
It satisfies the Mandelstam--Tamm (MT) relation~\cite{MT,F,B,V}
\be
 \arccos |\langle\psi(0)|\psi(T)\rangle| \le \int_0^T \diff t\,
 \sigma(H(t),|\psi(t)\rangle),
 \label{qsl0}
\ee
where $\sigma(H,|\psi\rangle)=\sqrt{
 \langle\psi|H^2|\psi\rangle
 -\langle\psi|H|\psi\rangle^2}$.
The left hand side of Eq.~(\ref{qsl0}) represents the Fubini--Study angle
and is used as a natural measure of the state separation.
It takes a positive value between 0 and $\pi/2$.
Equation (\ref{qsl0}) shows that 
the angle has a bound characterized by the energy variance.
Since the angle is interpreted as a distance measure, 
$\sigma$ plays a role of velocity.
This relation results from a general property of vectors in Hilbert space.
Applying a Hermitian operator $H$ to a state vector $|\psi\rangle$ gives
\be
 H|\psi\rangle = |\psi\rangle\langle \psi|H|\psi\rangle
 +|\psi_\perp\rangle\sigma(H,|\psi\rangle), \label{psiperp}
\ee
$|\psi_\perp\rangle$ is a normalized state orthogonal to $|\psi\rangle$.
The second term represents how the state deviates from the original one 
and is the origin of the bound in the MT relation.

{\it Quantum speed limit for adiabatic quantum computation.}
In the AQC, we are interested in obtaining 
the instantaneous ground state of the time-dependent Hamiltonian $H(t)$
by Schr\"odinger dynamics.
A slow driving approximately gives  
the adiabatic state $|\psi_{\rm ad}(t)\rangle$ 
whose formal definition is given in the following.
The performance of the computation is evaluated
by the fidelity between the ideal adiabatic state and
the time-evolved state $|\psi(t)\rangle$, 
\be
 \theta_{\rm ad}(t)=\arccos|\langle\psi_{\rm ad}(t)|\psi(t)\rangle|.
\ee
When we write the adiabatic state as a unitary time evolution
as $|\psi_{\rm ad}(t)\rangle=U_{\rm ad}(t)|\psi(0)\rangle$, the overlap is written
as $\langle\psi_{\rm ad}(t)|\psi(t)\rangle=\langle\psi(0)|\tilde{\psi}(t)\rangle$
where $|\tilde{\psi}(t)\rangle=U_{\rm ad}^\dag (t)|\psi(t)\rangle$.
The formal expression of the unitary operator $U_{\rm ad}(t)$ is given by
\be
 U_{\rm ad}(t)=\sum_n \e^{-i\int_0^t \diff t'\,\epsilon_n(t')
 -\int_0^t \diff t'\,\langle n(t')|\dot{n}(t')\rangle}
 |n(t)\rangle\langle n(0)|,  \label{u}
\ee
where $\{|n(t)\rangle\}$ represents a set of instantaneous eigenstates 
of $H(t)$ with the corresponding eigenvalues $\{\epsilon_n(t)\}$ and 
the dot denotes the time derivative.
The time derivative of $U_{\rm ad}(t)$ gives
\be
 i\frac{\diff U_{\rm ad}(t)}{\diff t}=(H(t)+H_{\rm CD}(t))U_{\rm ad}(t),
\ee
where
\be
 H_{\rm CD}(t)=i\sum_n (1-|n(t)\rangle\langle n(t)|)|
 \dot{n}(t)\rangle\langle n(t)|.
\ee
$H_{\rm CD}(t)$ is known as the counterdiabatic 
term in the method of 
shortcuts to adiabaticity (STA)~\cite{DR1,DR2,Berry,CRSCGM,STA,RMP}.
By adding this term to the original Hamiltonian, 
we can realize the adiabatic state of the original Hamiltonian
exactly by the time evolution.
Using this result, we find that
the generator of the state $|\tilde{\psi}(t)\rangle$
is given by 
$\tilde{H}(t)=-U_{\rm ad}^\dag(t)H_{\rm CD}(t)U_{\rm ad}^{\phantom \dag}(t)$.
Then, we can immediately apply the MT relation to obtain the bound  
\be
 && \theta_{\rm ad}(T) \le \int_0^T\diff t\,|\dot{\theta}_{\rm ad}(t)| \no\\
 && \le \min\left(
 \int_0^T \diff t\,\sigma(H_{\rm CD}(t),|\psi(t)\rangle), 
 \int_0^T \diff t\,\sigma(H_{\rm CD}(t),|\psi_{\rm ad}(t)\rangle)\right). 
 \no\\\label{qsl1}
\ee
The bound is characterized by two types of variance.
Since the counterdiabatic term is represented by using 
the time derivative of parameters in the original Hamiltonian, 
it is natural for this term to characterize the bound.
We can take the minimum of the variances to obtain a tight bound.
When the Hamiltonian $H(t)$ is prepared and we do not know 
the ideal adiabatic state $|\psi_{\rm ad}(t)\rangle$, the bound by 
the realistic state $|\psi(t)\rangle$ can be useful.
On the other hand, when we prepare $|\psi_{\rm ad}(t)\rangle$ 
as a reference state, 
the bound by $|\psi_{\rm ad}(t)\rangle$ would be appropriate.
The choice of the Hamiltonian $H(t)$ 
that corresponds to the specified state $|\psi_{\rm ad}(t)\rangle$ 
is not unique and we can obtain a universal bound 
which is common to all possible choices.
The variance with respect to $|\psi_{\rm ad}(t)\rangle$
has a geometrical meaning 
and appears when we discuss an energetic cost and
a trade-off relation for the implementation
of the counterdiabatic term~\cite{SS,CSHS,CD,FZCKUdC}.
In the following examples, we study bounds by 
$|\psi(t)\rangle$
since they give the same or slightly better results compared to those by  
$|\psi_{\rm ad}(t)\rangle$.

In the theory of QSL, we are basically interested in maximizing 
the left-hand side of Eq.~(\ref{qsl0}).
The MT relation shows that the maximum possible speed is 
given by the energy variance.
Here, we want to minimize $\theta_{\rm ad}(T)$, which means
that the speed limit, the rightmost side in Eq.~(\ref{qsl1}), 
gives a worst case evaluation of the performance.
Minimizing the variance can be an optimization method for the AQC.
In fact, the method of quantum adiabatic brachistochrone 
introduces a similar quantity 
for an optimization~\cite{RKHLZ,RALZ,Takahashi19}.

In the AQC, we expect that $\theta_{\rm ad}(t)$ is small and
$\dot{\theta}_{\rm ad}(t)$ oscillates around zero.
The original MT relation in Eq.~(\ref{qsl0})  
gives a tight bound only when $\arccos |\langle\psi(0)|\psi(t)\rangle|$
is a monotonic function. 
The same is true for $\theta_{\rm ad}(t)$ in Eq.~(\ref{qsl1}) 
and the equality is unlikely to hold in the present situation.
If we strictly impose the adiabaticity of the computation, 
the intermediate state $|\psi(t)\rangle$ at arbitrary $t$ 
is expected to be close to the adiabatic state $|\psi_{\rm ad}(t)\rangle$.
Basically, we are interested in the final state 
and it is not necessary for the intermediate state
to satisfy the adiabaticity.
However, we expect that an adiabatic-state following 
leads to a robust computation.
Then, $\int_0^T\diff t\,|\dot{\theta}_{\rm ad}(t)|$,
the middle term in Eq.~(\ref{qsl1}), becomes a
proper measure of adiabaticity and 
the rightmost side in Eq.~(\ref{qsl1}) can be a tight bound for 
this improved measure as we see in the following.

It is often a difficult task to calculate the explicit form of the bound.
The present result shows that
the bound is directly connected to the counterdiabatic term.
We know various ways to construct the counterdiabatic term
exactly~\cite{Jarzynski,delCampo,DJC,OT} 
and approximately~\cite{SP,OJV}, which would be useful to estimate
the bound.

{\it Some improvements.}
The bound can be improved by imposing additional conditions.
One of the simplest conditions is to set that the initial state
$|\psi(0)\rangle$ is one of the eigenstates
of the initial Hamiltonian $H(0)$, $|n(0)\rangle$.
This is a natural condition usually employed in the AQC.
In this case, the adiabatic state is written 
by the single eigenstate $|n(t)\rangle$.
The dynamical phase factor $\e^{-i\int_0^t \diff t'\,\epsilon_n(t')}$
gives an overall contribution and is dropped out
when we take the absolute value of the overlap.
The time evolution is effectively achieved only by the counterdiabatic term,
that means $U_{\rm ad}(t)$ is equivalent to
$\e^{-i\int_0^t \diff t'\,\epsilon_n(t')}U_{\rm CD}(t)$ where 
$U_{\rm CD}(t)$ is the time-evolution operator for $H_{\rm CD}(t)$ and
is obtained by setting $\epsilon_n(t)=0$ for $U_{\rm ad}(t)$ in Eq.~(\ref{u}). 
We find that the bound is obtained by replacing 
$\sigma (H_{\rm CD}(t),|\psi(t)\rangle)$ 
in Eq.~(\ref{qsl1}) by 
$\sigma (H(t)-H_{\rm CD}(t),|\psi(t)\rangle)$. 
Although this bound is expected to be an improvement over that in 
Eq.~(\ref{qsl1}),
it is not evident whether $\sigma (H(t)-H_{\rm CD}(t),|\psi(t)\rangle)$ 
is smaller than $\sigma (H_{\rm CD}(t),|\psi(t)\rangle)$.
To obtain some intuition, we find a further different expression 
in the following.

The idea also comes from the method of STA.
In the counterdiabatic driving, we introduce the additional
counterdiabatic term $H_{\rm CD}(t)$ to the original Hamiltonian $H(t)$
to keep the adiabatic state with respect to $H(t)$.
The idea of STA is not restricted to this procedure 
and we can consider several variants of implementations.
In fact, in the ``inverse engineering'', we use the dynamical invariant
operator to obtain an ideal time evolution~\cite{CRSCGM,LR}.
From a viewpoint of the counterdiabatic driving, 
the use of the dynamical invariant corresponds to 
decomposing the Hamiltonian into two parts: $H(t)=H_0(t)+H_1(t)$.
The first term $H_0(t)$ commutes with the dynamical invariant.
The solution of the Scr\"odinger equation with $H(t)$ is given by the
adiabatic state of $H_0(t)$, which means that
$H_1(t)$ is interpreted as the counterdiabatic term.
$H_1(t)$ is different from $H_{\rm CD}(t)$ as we discuss below.
When we start the time evolution from an eigenstate 
of the initial Hamiltonian, we determine the decomposition as follows. 
We prepare the basis $\{U(t)|n(0)\rangle\}$, 
where $U(t)$ is the time-evolution operator for $H(t)$.
This basis represents a set of eigenstates for the dynamical invariant.
Then, $H_0(t)$ represents the diagonal part 
and $H_1(t)$ the offdiagonal part~\cite{Takahashi17}.
Setting the initial state as $|\psi(0)\rangle=|n(0)\rangle$ and 
following the same logic as before, 
we see that $|\psi(t)\rangle$ is obtained,
up to the phase, by applying the unitary operator $U_1(t)$ 
for $H_1(t)$ to the initial state.
Then, we obtain the bound with 
$\sigma (H_1(t)-H_{\rm CD}(t),|\psi(t)\rangle)$ 
in place of $\sigma (H(t)-H_{\rm CD}(t),|\psi(t)\rangle)$.

{\it Adiabatic expansion.} 
The representation of the bound using the difference between two
counterdiabatic terms is instructive.
Although the explicit operator form of $H_1(t)$ is generally hard to obtain, 
we can find an intuitive meaning by using the adiabatic expansion.
The dynamical invariant operator $F(t)$ satisfies 
\be
 i\frac{\diff F(t)}{\diff t}=[H(t),F(t)],
\ee
and is solved formally by using the expansion in terms of
the time derivative operator.
Since the dynamical invariant commutes with $H_0(t)$,
$H_1(t)$ is obtained by solving the commutation relation
$[H(t)-H_1(t),F(t)]=0$.
Solving these equations order by order, we find 
\be
 &&\langle m(t)|H_1(t)|n(t)\rangle\no\\
 && = \langle m(t)|H_{\rm CD}(t)|n(t)\rangle
 +i\frac{\diff}{\diff t}\frac{\langle m(t)|H_{\rm CD}(t)|n(t)\rangle}
 {\epsilon_m(t)-\epsilon_n(t)}+\cdots, \label{adexp}
\ee
for $m\ne n$.
The diagonal part $\langle n(t)|H_1(t)|n(t)\rangle$ 
is not required when we calculate the variance.
The result shows that $H_1(t)$ is equivalent to $H_{\rm CD}(t)$
at first order of the expansion.
We use the variance of $H_1(t)-H_{\rm CD}(t)$ for the bound,
which means that the bound can be approximately characterized by
the second term of Eq.~(\ref{adexp}) in the adiabatic regime.
We obtain
\be
 \sigma^2 (H_1(t)-H_{\rm CD}(t),|\psi(t)\rangle)
 \sim\sum_{m(\ne n)} \left|
 \frac{\diff}{\diff t}\left(
 \frac{\langle m(t)|H_{\rm CD}(t)|n(t)\rangle}
 {\epsilon_m(t)-\epsilon_n(t)}\right)\right|^2. \no\\
\ee
Since $H_{\rm CD}(t)$ incorporates the time derivative of the parameters,
the second time derivative of the parameters is relevant, 
rather than the first derivative, 
for characterizing the bound of the fidelity.
The acceleration can be relevant for some control problems.
In the counterdiabatic driving, the acceleration potential
is obtained by using a unitary transformation~\cite{Takahashi15}.
The relevance of the higher-order derivatives
for the adiabatic approximation can also be seen in rigorous treatments
of the adiabatic theorem~\cite{JRS,LRH,BS,EH,AL18}.

{\it Two-level systems.}
We study a simple two-level system to see how tight the obtained bounds are. 
The Hamiltonian is given by 
$H(t)=\frac{h}{2}(\sigma^z\cos\theta(t)+\sigma^x\sin\theta(t))$, 
where $\sigma^z$ ($\sigma^x$) is the $z$ ($x$) component
of the Pauli matrices.
$h$ is fixed to a constant value and 
$\theta(t)$ moves from $\theta(0)=\pi/2$ to $\theta(T)=0$.
In this case, $H_{\rm CD}(t)$ is given by 
$H_{\rm CD}(t)=\frac{\dot{\theta}(t)}{2}\sigma^y$.
We also find 
$\sigma (H(t)-H_{\rm CD}(t),|\psi(t)\rangle)=\sigma (H_1(t)-H_{\rm CD}(t),|\psi(t)\rangle)$.
Then, we compare $\Delta E_1(t)=\sigma (H_{\rm CD}(t),|\psi(t)\rangle)$
and $\Delta E_2(t)=\sigma (H(t)-H_{\rm CD}(t),|\psi(t)\rangle)$
as possible bounds.

\begin{center}
\begin{figure}[t]
\begin{center}
\includegraphics[width=1.\columnwidth]{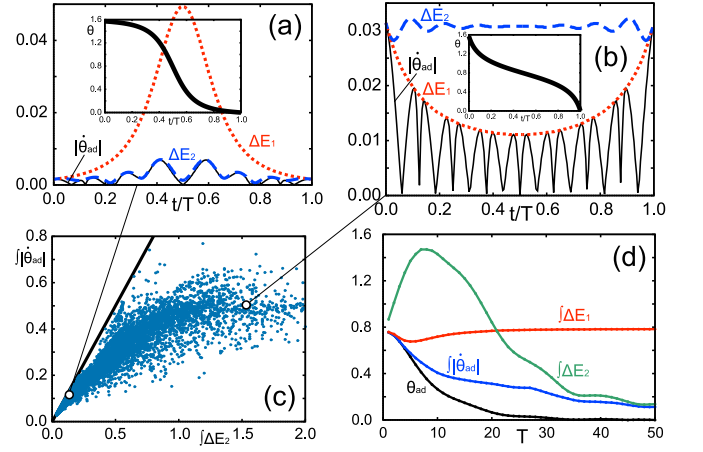}
\end{center}
\caption{
QSL for a two-level system.
(a), (b) 
$|\dot{\theta}_{\rm ad}(t)|$ (black solid),
$\Delta E_1(t)$ (red dotted), and $\Delta E_2(t)$ (blue dashed) 
for a protocol $\theta(t)$ given in the inset of each panel.
We set the annealing time $T=50$ in units of $h$.
(c) Distribution of
$\left(\int_0^T \diff t\,\Delta E_2(t), \int_0^T \diff t\,|\dot{\theta}_{\rm ad}(t)|\right)$
for randomly generated protocols $\theta(t)$.
The protocols are generated as decreasing functions from $\pi/2$ to 0. 
The solid line represents the bound.
(d) Annealing time dependence for 
the protocol in the top left-hand panel.
$\dot{\theta}_{\rm ad}(t)$ (black),
$|\dot{\theta}_{\rm ad}(t)|$ (blue),
$\Delta E_1(t)$ (red), and 
$\Delta E_2(t)$ (green)
are integrated from 0 to $T$.
}
\label{fig-two}
\end{figure}
\end{center}

The numerical study is summarized in Fig.~\ref{fig-two}.
$|\dot{\theta}_{\rm ad}(t)|$ becomes small
when the parameter changes around the initial and final times are slow,
as we see in the top left-hand panel of Fig.~\ref{fig-two}.
In this case, $\Delta E_2(t)$ gives a good tight bound.
The importance of the slow changes at the boundaries
has been discussed in several works~\cite{Morita,LRH,AL15}.
Our observation is consistent with their results.
It should be remarked that we can find a good bound
irrespective of the performance.
Although the oscillations of $|\dot{\theta}_{\rm ad}(t)|$
in the top right-hand panel are difficult to be captured by the bound, 
the bound by $\Delta E_1(t)$ can describe the outline of 
the oscillation as an envelope.

The annealing-time dependence of the result is shown in 
the bottom right-hand panel of Fig.~\ref{fig-two}.
$\Delta E_2(t)$ is strongly dependent on the annealing time $T$
while $\Delta E_1(t)$ is not sensitive to $T$.
$\Delta E_2(t)$ becomes small at large $T$,
which is understood from the adiabatic expansion as
discussed in Eq.~(\ref{adexp}).

\begin{center}
\begin{figure}[t]
\begin{center}
\includegraphics[width=1.\columnwidth]{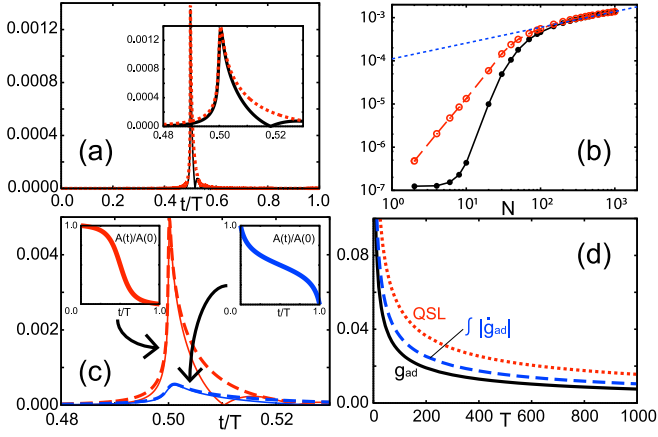}
\end{center}
\caption{
QSL for a spin chain model.
(a) $|\dot{g}_{\rm ad}(t)|$ (black solid line)
and the right hand side of Eq.~(\ref{qslmb2}) (red dotted)
for a linear protocol $A(t)=A(0)(1-t/T)$, $B(t)=A(0)t/T$.
We set $N=1000$ and $T=1000$ in units of $A(0)$.
The inset represents a blowup around the peak.
(b) Size dependence of the peak height obtained in (a).
The black solid line with the symbol $\bullet$
denotes $\max_t|\dot{g}_{\rm ad}(t)|$ and 
the red dashed line with $\circ$ the corresponding quantity in the QSL.
They show the same power-law behavior 
denoted by the blue dotted line at large $N$. 
(c) Protocol dependence.
We choose two types of $A(t)$ shown in the insets 
and $B(t)=A(0)-A(t)$.
The solid lines denote $|\dot{g}_{\rm ad}(t)|$ and 
the dashed lines the QSL.
We set $N=1000$ and $T=1000$ in units of $A(0)$.
(d) Annealing time dependence. 
We use the linear protocol used in (a) and set $N=200$.
}
\label{fig-1d}
\end{figure}
\end{center}

{\it Quantum speed limits for many-body systems.}
In the AQC, our interest is mainly on systems with many degrees of freedom.
Although the obtained bounds are applicable to any closed quantum systems,
they are not useful in typical many-body systems.
For a system with the particle number $N$, the Hamiltonian is
an extensive quantity and 
the state is basically given by a product of $N$ components.
This means that the fidelity is expected to have a form 
\be
 \langle\psi(0)|\psi(t)\rangle\sim \e^{-Ng(t)}, 
\ee
where $g(t)$ is a non-negative function independent of $N$.
This becomes a very small quantity for a large $N$.
In other words, the size of the Hilbert space is too huge 
for two vectors to have a certain amount of overlap.
The vanishing of the overlap can be found 
even when we consider a small perturbation.
It is called the orthogonality catastrophe~\cite{Anderson} and 
has recently been discussed from a viewpoint of the QSL~\cite{FDBC}. 
The behavior of the fidelity
for many-body systems can be studied by using the rate function $g(t)$.
In fact, a dynamical singularity appears on this quantity for 
systems with quantum quench~\cite{HPK}.
When the rate function becomes a well-defined quantity,
the overlap immediately goes to zero at $N\to\infty$.
Since $\sigma(H(t),|\psi(t)\rangle)$ is typically proportional to $\sqrt{N}$, 
the MT relation becomes a trivial one.

To find a meaningful relation, we reexamine the derivation of the MT relation.
Using Eq.~(\ref{psiperp}), we find  
\be
 |\dot{g}(t)| \le \frac{\sigma (H(t),|\psi(t)\rangle)}{N}\left|
 \frac{\langle\psi(0)|\psi_\perp(t)\rangle}{\langle\psi(0)|\psi(t)\rangle}
 \right|
 =\frac{\sigma (H(t),|\psi(t)\rangle)}{\sqrt{N}}
 \frac{c_\perp(t)}{c(t)}. \no\\ \label{qslmb}
\ee
The equality is obtained when 
the ratio $\langle\psi(0)|\psi_\perp(t)\rangle/\langle\psi(0)|\psi(t)\rangle$
becomes pure imaginary.
When the fidelity is scaled as 
$|\langle\psi(0)|\psi(t)\rangle|= c(t)\e^{-Ng(t)}$ where $c(t)$ and $g(t)$ are 
non-negative and independent of $N$, we see below
$|\langle\psi(0)|\psi_\perp(t)\rangle|= \sqrt{N}c_\perp(t)\e^{-Ng(t)}$, 
with $c_\perp(t)$ which is also non-negative and independent of $N$.
As a result, we obtain the last expression in Eq.~(\ref{qslmb}).
The right-hand side remains finite even if we take the thermodynamic
limit $N\to\infty$.
Then, we can use this relation as a new type of QSL for many-body systems.
We note that this inequality makes sense even for small systems.
Since the present relation does not require 
an additional inequality
$|\langle \psi(0)|\psi(t)\rangle|^2+|\langle \psi(0)|\psi_\perp(t)\rangle|^2\le 1$
which is used to derive the MT relation,
we expect that Eq.~(\ref{qslmb}) gives a tighter bound.

It is not convenient to represent the bound by using 
the unknown state $|\psi_\perp(t)\rangle$.
The bound can be represented by the counterdiabatic term.
Setting the condition that the initial state is in one of the eigenstate 
of the initial Hamiltonian, we obtain 
$H_1(t)|\psi(t)\rangle=|\psi_\perp(t)\rangle\sigma (H(t),|\psi(t)\rangle)$ and 
\be
 |\dot{g}(t)|\le \frac{1}{N}\left|
 \frac{\langle\psi(0)|H_1(t)|\psi(t)\rangle}{\langle\psi(0)|\psi(t)\rangle}
 \right|. \label{qslmb1}
\ee
Since the counterdiabatic term is expected to be an extensive operator,
we see that the right-hand side remains finite
even if we take the limit $N\to\infty$.
It is interesting to see that the quantity appearing  
on the right-hand side represents the weak value~\cite{AAV}.

In a similar way, for the fidelity with the adiabatic state,
we define $g_{\rm ad}(t)$ as 
$|\langle\psi_{\rm ad}(t)|\psi(t)\rangle|\sim\e^{-Ng_{\rm ad}(t)}$ 
to derive the bound 
\be
 |\dot{g}_{\rm ad}(t)|\le \frac{1}{N}\left|
 \frac{\langle\psi_{\rm ad}(t)|H_{\rm CD}(t)|\psi(t)\rangle}
 {\langle\psi_{\rm ad}(t)|\psi(t)\rangle}
 \right|. \label{qslmb2}
\ee
With an additional condition that 
the initial state is in one of the eigenstate of the initial Hamiltonian, 
we can replace $H_{\rm CD}(t)$ in Eq.~(\ref{qslmb2}) by
$H(t)-H_{\rm CD}(t)$ and $H_1(t)-H_{\rm CD}(t)$
as we have shown in the previous calculations.

\begin{center}
\begin{figure}[t]
\begin{center}
\includegraphics[width=1.\columnwidth]{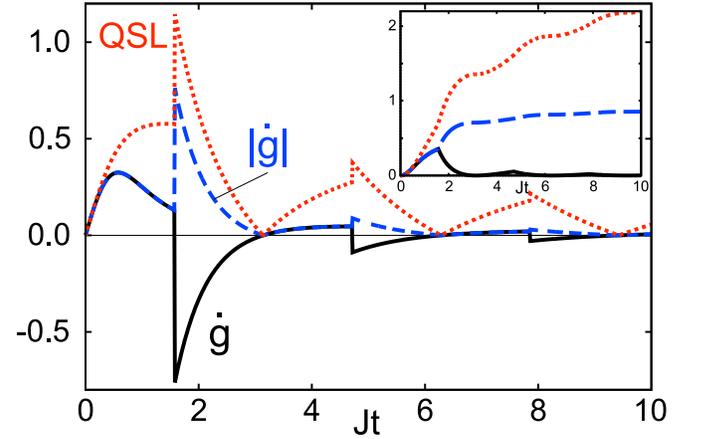}
\end{center}
\caption{QSL for a quantum quench system with dynamical phase transitions.
We set $J=h$.
The black solid line represents $\dot{g}(t)$, 
the blue dashed line $|\dot{g}(t)|$, and the red dotted line
the QSL specified by the right-hand side in Eq.~(\ref{qslmb1}).
Inset: Time-integrated quantities.
}
\label{fig-quench}
\end{figure}
\end{center}

{\it Some examples.}
We study many-body spin models that exhibit phase transitions.
First, we treat the transverse-field Ising-spin chain, 
\be
 H(t)= -\frac{A(t)}{2}\sum_{i=1}^N\sigma_i^x
 -\frac{B(t)}{2}\sum_{i=1}^{N}\sigma_i^z\sigma_{i+1}^z.
\ee
We use the periodic boundary condition  $\sigma_{N+1}^z=\sigma_1^z$.
This Hamiltonian can be decomposed into a set of two-level systems
by the Jordan--Wigner transformation~\cite{JW, LSM}.
The result is shown in Fig.~\ref{fig-1d}.
The QSL represented by Eq.~(\ref{qslmb2}) gives a tight bound
even at the quantum phase-transition point $A=B$
obtained by the static treatment. 
We observe a peak at the point and 
the height is scaled by the size of the system as 
$N^\alpha$ with $\alpha\sim 0.303$
in the present choice of parameters.
The same scaling is applied to both the fidelity and the QSL, which implies
that the universal properties at the phase transition can be 
studied by using the QSL.
 
A different type of singularity can be
found for a quantum quench system 
and is known as dynamical phase transitions~\cite{HPK}.
We consider the spin operator $\bm{S}$ with 
$\bm{S}^2=\frac{N}{2}\left(\frac{N}{2}+1\right)$, 
and prepare the initial state $|\psi(0)\rangle$
as an eigenstate of $S^x$, 
$S^x|\psi(0)\rangle=\frac{N}{2}|\psi(0)\rangle$.
Then, the state is time-evolved under the Hamiltonian 
\be
 H=-2\left(\frac{J}{N}(S^z)^2+hS^z\right).
\ee
It is known that the rate function $g(t)$ at $N\to \infty$
has singular points~\cite{OST}.
The decomposition of the Hamiltonian is possible in this case~\cite{Takahashi17}
and we can calculate the bound as we show in Fig.~\ref{fig-quench}.
$\dot{g}(t)$ changes discontinuously at phase-transition points.
We find that the QSL still holds even when we have the transitions.
We also see that the bound can basically be a good estimate of
the rate function but 
it becomes loose around the transition points.
Since $H_1(t)$ is a part of the original Hamiltonian, the bound 
in Eq.~(\ref{qslmb1}) stays finite, which indicates that 
$g(t)$ does not diverge in any systems.

{\it Conclusion.} 
We have discussed the QSL applied to the AQC.
The performance of the computation 
is characterized by the counterdiabatic term.
The bound is simply represented by the variance of 
the counterdiabatic term and has a geometrical meaning.
Although we mainly focused on the AQC, 
the result is general enough so that it can be used 
without any additional condition.
As we mentioned in the Introduction, the present method
makes up for the inconvenience of the previous methods.
We also find a novel type of QSL
that can be applied to many-body systems.
Our result implies that the universal properties can be deduced
from the corresponding counterdiabatic term.


We are grateful to Adolfo del Campo, Ken Funo, Takuya Hatomura, and Yutaka Shikano
for useful discussions and comments.


\end{document}